# EXACTLY SOLVABLE MODELS : A SOLUTION FOR DIFFERENT PROBLEMS OF LASER MATTER INTERACTION

*By* : Guillaume PETITE [1] and Alexander B. SHVARTSBURG [2]

(1) Laboratoire des Solides Irradiés, CEA/DSM, CNRS (UMR 7642) and Ecole Polytechnique, 91128, Palaiseau CEDEX, France. *tel* : 33-1 69 33 44 96, *fax* : 33-1 69 33 30 22, email :guillaume.petite@polytechnique.fr

(2) Central Design Bureau for Unique Instrumentation of the R A S, Butlerov Str. 15, Moscow, Russia
*tel*: 095 334 83 49, *fax*: 095 334 75 00, email:

**ABSTRACT**

With the increasing use of ultrashort laser pulses and nanoscale-materials, one is regularly confronted to situations in which the properties of the media supporting propagation are not varying slowly with time (or space). Hence, the usual WKB-type approximations fail, and one has to resort to numerical treatments of the problems, with a considerable loss in our insight into the physics of laser-matter interaction. We will present a new approach which allows a fully analytical solution of such problems, based on a transformation of the propagation equations into a new space where phase accumulation is linear with either time or space, which greatly simplifies their treatment. Though this method is restricted to some special models of the time or space varying dielectric constant, those are however general enough to encompass practically all experimental situations. It allows to introduce the concept of "non-stationarity induced (or "inhomogeneity induced") dispersion. We will analyse the problem of reflection and propagation in two types of media whose dielectric constant vary rapidly at either the laser period or the laser wavelength scale. Extension of such techniques to the case of arbitrarily high non linearities will be considered too.

## 1. Introduction

The increasing use in the past few years of both nanoscale optical materials and of femtosecond laser pulses, including ultra high intensity ones, has brought back to the forefront a number of problems which, if not overlooked, were not considered as crucial ones. Whether one considers the space-like or the time-like problems, it was a quite common attitude to keep working in the frame of the so called "slowly varying envelope approximation" (or any other WKB-type one) even though acknowledging the fact that one was close to its limits, if not beyond. One now faces so often cases where substantial variations ( of the order of 100%) of the material properties (that will take the form of a space- or time-dependent dielectric constant) over length scales well below the wavelength or times scales of the order of the light period that time has come to seek new approaches in order to investigate such situations keeping as much as we can of our insight into physics, which fully numerical treatments (another tentative solution to such problems) do not do best. Two typical examples concern the use of the so-called "index gradient " layers[1,2], in which the refractive index can vary by as much as 60% over typically 100 nm (space-like case), or the reflection of a high intensity laser pulse on a typical optical material like $SiO_2$, where due to free-carrier injection, the material can transit from a transparent state to a quasi-metallic one (except for details about the degenerate character of the electron distribution) in one or a few laser periods (time-like case)[3].

In this paper, we will describe a new method by which it is possible to tackle such problems by completely analytical methods : with a varying dielectric constant, phase accumulation is no longer linear with space (or time), with the consequence that the propagation equation becomes unsolvable. Therefore, we propose to restore this linearity through a transformation of the space in which the equations are written. By doing so, one recovers analytical simplicity, and the solution of the propagation equation in the new space so defined may even, under some restrictions, express under the classical form of progressive sinusoidal waves. It is important that the above mentioned restrictions do not affect the generality of the types of variation of the dielectric constant we can consider (monotonous or not, increasing or decreasing...), which happens to be the case. This method, to which we will refer as to the "phase coordinate method", keeps unharmed our physical insight into the physics of propagation, allowing us to reveal a new type of dispersion induced by the material inhomogeneity (depending directly on the gradients of the dielectric constant rather than on its local value) or non stationarity. In the following, we will exemplify this method through the investigation of two types of media, namely with space and time varying dielectric constant, of which the reflection properties will be studied.

Finally, we will show that at the expense of a "hodagraph" transform (by which the time and space become the variables depending on electric and magnetic fields as coordinates) the problem of an arbitrary high non linearity can be cast under an equivalent analytical form as the two problems above, and can thus be treated equivalently.

## 2. Inhomogeneous media : " index gradient " nanolayers

Let us consider an inhomogeneous dielectric film as a plane dielectric layer with thickness $d$ and dielectric susceptibility $\varepsilon(z)$, $0 \leq z \leq d$. This susceptibility will be cast under the form :

$$\varepsilon(z) = n_0^2 U^2(z) \tag{1}$$

A linearly polarized EM wave, that we assume to propagate in the z-direction (normal incidence conditions), is described by Maxwell equations, linking the $E_x$ and $H_y$ components of the wave :

$$\frac{\partial E_x}{\partial z} = -\frac{1}{c}\frac{\partial H_y}{\partial t}$$

$$\frac{\partial H_y}{\partial z} = -\frac{\varepsilon(z)}{c}\frac{\partial E_x}{\partial t} \tag{2}$$

The EM field in normal incidence can be described with help of a single component vector-potential cast under the form $\boldsymbol{A}(z,t) = \boldsymbol{A_0}\,\psi(z,t)$, where $\psi$ is a scalar function such that (for convenience, we put $A_0=1$)

$$E_x = -\frac{1}{c}\frac{\partial \psi}{\partial t} \;;\; H_y = \frac{\partial \psi}{\partial z} \tag{3}$$

which allows to reduce the system (1)-(2) to the single equation

$$\frac{\partial^2 \psi}{\partial z^2} - \frac{n_0^2 U^2(z)}{c^2}\frac{\partial^2 \psi}{\partial t^2} = 0 \tag{4}$$

Clearly, if $U$ was constant this would be the traditional propagation equation in an optically transparent medium, admitting the well known propagating wave solutions. In general, equation (4) with a space varying $U(z)$ is unsolvable. However, using a new function $F$ and a new variable $\eta$

$$F = \psi\sqrt{U(z)} \;;\; \eta = \int_0^z U(\zeta)d\zeta \tag{5}$$

transforms eq.(4) into a new equation

$$\frac{\partial^2 F}{\partial \eta^2} - \frac{n_0^2}{c^2}\frac{\partial^2 F}{\partial t^2} = \frac{F}{2U^3(z)}\left[\frac{\partial^2 U}{\partial z^2} - \frac{3}{2U(z)}\left(\frac{\partial U}{\partial z}\right)^2\right] \qquad (6)$$

We will use the following model of dielectric susceptibility profile $\varepsilon(z)$

$$\varepsilon(z) = n_0^2 U^2(z) \text{ , with } U(z) = \left(1 + \frac{s_1 z}{L_1} + \frac{s_2 z^2}{L_2^2}\right)^{-1} \text{ ; } s_1 = 0, \pm 1 \text{ ; } s_2 = 0, \pm 1 \qquad (7)$$

Here $n_0$ is the refractive index value on the interface z = 0; the distribution (7) is considered in the region z ≥ 0. The characteristic spatial scales $L_1$ and $L_2$ as well as the values $s_1$ and $s_2$ are the free parameters of model (7). Note that the well known Rayleigh profile – $U(z)=Az^{-2}$, the only case where a general solution of eq (4) was known - corresponds to the limit of $U(t)$ when the scale $L_2 \to \infty$, so that we are in fact generalizing this profile, with the important result that practically all variations of the dielectric can be at least qualitatively represented with such a model, due to the increased number of free parameters. With such a model, eq (6) now has constant coefficients, writing :

$$\frac{\partial^2 F}{\partial \eta^2} - \frac{n_0^2}{c^2}\frac{\partial^2 F}{\partial t^2} = p^2 F \qquad (8)$$

with

$$p^2 = \frac{1}{4L_1^2} - \frac{s_2}{L_2^2} \qquad (9)$$

A solution of eq.(8) can be built by superposition of waves with wavenumber q, travelling in the $\eta$ - direction

$$F = \exp i(q\eta - \omega t) \qquad (10)$$

$$q = \frac{\omega n_0}{c} N \text{ ; } N = \sqrt{1 - \frac{c^2 p^2}{n_0^2 \omega^2}} \qquad (11)$$

We note that solutions in the form of travelling waves are obtained only if the expression under the radical is positive. This is always the case if $p^2<0$, i.e. for $s_2=+1$, if $L_2>2L_1$. In the opposite case ($p^2>0$), the availability of travelling wave solutions is subject to a condition concerning $\omega$, which writes

$$\omega > \Omega \text{ with } \Omega = \frac{c\sqrt{y^2 - s_2}}{n_0 L_2} \text{ and } y = \frac{L_2}{2L_1} \qquad (12)$$

so that such films possess a cut-off frequency. Combining (10) and (5), we obtain the function $\psi$ determining the vector-potential; whose substitution into (4) brings the explicit expressions for the field components

$$E_x = \frac{i\omega}{c\sqrt{U}} \exp i(q\eta - \omega\tau) \qquad (13)$$

$$H_y = \frac{i\omega n_0 \sqrt{U}}{c}(N - iG)\exp i(q\eta - \omega\tau) \qquad (14)$$

with

$$G = -\frac{c}{2\omega n_0 L_1}\left(1 + \frac{2s_2 z L_1}{s_1 L_2^2}\right) \qquad (15)$$

Thus we found an exact solution describing the EM wave in an inhomogeneous layer (7). At this point, let us consider the dispersion relation $q(\omega)$ given in eq. 10. It can be described as "waveguide-like" or "plasma-like", the equivalent "plasma frequency" being the quantity $cp/n_0$. Now it is essential to remark the parameter $p$ depends only on the characteristic lengthscales of the layer's index variation. This thus allows to introduce the concept of "inhomogeneity-induced" dispersion, which has nothing to do with the "natural" (local) dispersion of the material (contained, eventually in $n_0$). The same can be said about the cut-off frequency, when it exists.

Another comment concerns the "phase coordinate" which has here a quite simple meaning: from the definition (5), it is obvious that it is simply the optical path.

From the field expressions in the inhomogeneous material one can, using the standard method of the field continuity at the interfaces, calculate the reflection coefficient of a wave on such a layer deposited on a substrate. The calculations are not specially difficult, though somewhat heavy, and we just give here the essential and typical results, the reader being referred to the original publication[4] for their complete derivation. We consider first the case of a film presenting a cut-off frequency $\Omega_1$ ($p^2>0$, here $s_2<0$). Figure 1a shows the reflection coefficient of such a film "stand alone" as a function of the scaled frequency $\omega/\Omega_1$. One sees that such a film has antireflection properties which extend over a considerable frequency range. Figure 1b shows the reflection coefficient of such a film deposited on an absorbing substrate with a complex refractive index equal to $3.5+i\,0.7$, which would otherwise have an intensity reflection coefficient of 0.32.

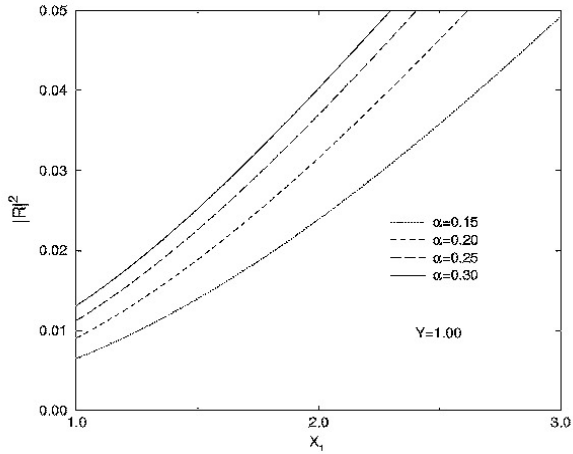 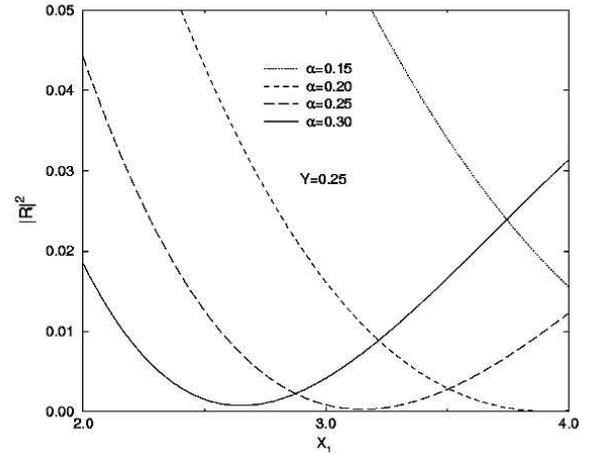

Figure 1,a: Antireflection properties of thin inhomogeneous dielectric film in the case $s_2 <0$ and $p^2 > 0$. The reflection coefficient $|R|^2$ is plotted vs the normalized frequency $x_1 = \omega/\Omega_1$ for $n_0 = 1.73$, and for different values of the parameter $\alpha = d/L_2$.

Figure 3 : Broadband antireflection properties of a thin inhomogeneous film, coating a lossy dielectric with $n_2 = 3.5$, $\chi_2 = 0.7$; the dependence of the reflectivity $|R|^2$ upon the normalized frequency $x_1 = \omega/\Omega_1$, relating to the case $y = 0.25$, $n_0 = 1.6$, $p^2>0$, $s_2<0$, is represented for different values of the parameter $\alpha = d/L_2$. The same dielectric without coating has a reflection coefficient $|R|^2 = 0.32$

Let us note here that the scaling frequency depends only on geometrical parameters so that, by changing the geometric characteristics of the coating (length scales of the inhomogeneity) one could achieve antireflection coatings of this type in almost any wavelength range (of course there will be a natural limitation in the XUV due to the only assumption here, which is that the film index varies continuously). Other films can show a quite different behavior, as shown on figure 2, which presents the case of a "stand alone" film with $p^2 < 0$ and $s_2 > 0$. Such films do not posses a cut-off frequency, and the scaling frequency is in this case defined as :

$$\Omega_2 = \frac{c\sqrt{1-y^2}}{n_0 L_2} \qquad (16)$$

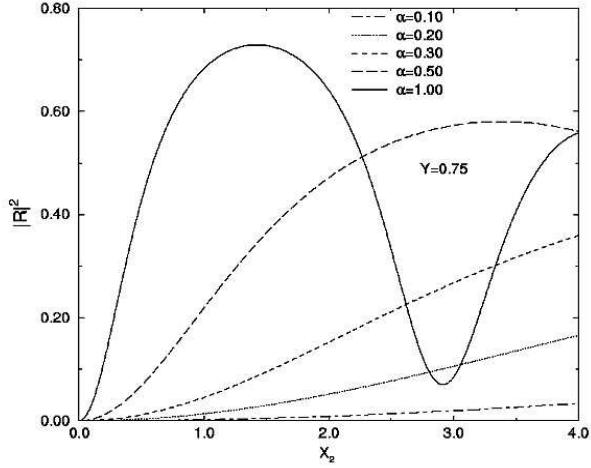

Figure 2 : Spectra of reflectivity of inhomogeneous film in a case $s_2>0$ and $p^2 < 0$; The reflection coefficient $|R|^2$ is plotted vs. the normalized frequency $x_2 = \omega/\Omega_2$ for $n_0 = 1.73$, and for different values of the parameter $\alpha = d/L_2$. The inhomogeneity induced frequency $\Omega_2$ is given in (29).

One sees that such films have a quite different behavior from those described above. In particular (as it can be clearly seen in the case of $\alpha=1$ on fig 2, such films have a dichroic character.

To summarize this first application of our method to the case of a space dependent dielectric constant, transforming the wave equation into a space where the optical path is the space-like variable we can, under some restriction concerning the type of variation – which do not preclude the possibility of encompassing many different types of dependencies - obtain analytical solutions for the fields in this space under the form of progressive waves, from which the behavior in the normal space can be recovered, and allowing to compute different properties such as the reflection coefficients of such films. We find that they present a plasma-like dispersion law, which is completely determined by the geometrical constants defining the space dependence of the media dielectric constant. Depending of such parameters, the film can present a cut-off frequency of not and different reflection properties.

## 3. Non-stationary media

Let us now consider a case in which the dielectric constant depends on time instead of space. The same method can be applied in the following way[5]: we consider here the case of transverse electric wave ($z$ being the propagation direction). One starts from the Maxwell equations for such a case :

$$\frac{\partial E_x}{\partial z} = -\frac{1}{c}\frac{\partial H_y}{\partial t}; \frac{\partial E_x}{\partial y} = \frac{1}{c}\frac{\partial H_z}{\partial t}; \frac{\partial H_z}{\partial y} - \frac{\partial H_y}{\partial z} = \frac{1}{c}\frac{\partial D_x}{\partial t} \qquad (17)$$

using the generating function

$$E_x = -\frac{1}{c}\frac{\partial \psi_s}{\partial t}; H_y = \frac{\partial \psi_s}{\partial z}; H_z = -\frac{\partial \psi_s}{\partial y} \qquad (18)$$

one obtains the following wave equation

$$\frac{\partial^2 \psi_S}{\partial z^2} + \frac{\partial^2 \psi_S}{\partial y^2} - \frac{n_0^2 U^2(t)}{c^2}\frac{\partial^2 \psi_S}{\partial t^2} = \frac{n_0^2}{c^2}\frac{\partial U^2}{\partial t}\frac{\partial \psi_S}{\partial t} \qquad (19)$$

which is also not solvable as this. We then make the following transformation :

$$F_S = \psi_S \sqrt{U(t)} \; ; \; \tau = \int_0^t \frac{dt_1}{U(t_1)} \tag{20}$$

by which the wave equation becomes

$$\frac{\partial^2 F_S}{\partial z^2} + \frac{\partial^2 F_S}{\partial y^2} - \frac{n_0^2}{c^2}\frac{\partial^2 F_S}{\partial \tau^2} = -\frac{n_0^2}{c^2} F_S \left( \frac{UU_{tt}}{2} + \frac{U_t^2}{4} \right) = \frac{n_0^2}{c^2} F_S G \tag{21}$$

($U_t$ and $U_{tt}$ represent the first and second time-derivatives of $U$). Once again, with some restriction on the analytical form of $U$, which is still flexible enough to account for any typical variation $\varepsilon(t)$ (monotonous or not, increasing or decreasing..), plane wave solutions to the propagation equation in this space are available :

$$F_S = \exp i(k_\perp z + k_{II} y - \omega \tau)$$
$$k^2 = k_\perp^2 + k_{II}^2 = \left(\frac{\omega n_0}{c}\right)^2 N_S^2 ; \tag{22}$$
$$N_S^2 = 1 - \frac{p_S^2}{\omega^2}$$

where $p_s^2$ is here determined only by the time constants of the function $U$, obtained here in the $\tau$-space as (depending on the sign of $p_s^2$) :

$$p_S^2 = p_1^2 = -T_1^{-2} < 0 :$$
$$U_1(\tau) = Q_1^2 = \left[ \cos(\tau/T_1) + M_1 \sin(\tau/T_1) \right]^2 \tag{23}$$

or

$$p_S^2 = p_2^2 = T_2^{-2} > 0$$
$$U_2(\tau) = Q_2^2 = \left[ ch(\tau/T_2) + M_2 sh(\tau/T_2) \right]^2 \tag{24}$$

The case of a transverse magnetic wave can be treated in a similar way. It should be noted that the analytical dependencies of $U(t)$ obtained in both cases are different, but that once fixed the essential parameters (amplitude of the variation and characteristic time) the difference between both functions are negligible[5]. Now, looking at eqs. (22), one sees that the same remark as in the space-dependent case can be made : the dispersion law for the wave in the non-stationary medium is plasma-like, and is determined by the time constants of the dielectric function's variations (we will speak here of a "non-stationarity induced dispersion").

The fields in the medium can be calculated from eqs (18), (20) and (22) :

$$\psi_S = \frac{A_S \exp i(k_\perp z + k_{II} y - \omega \tau)}{\sqrt{U(t)}} \tag{25}$$

$$E_x = \frac{i\omega}{cU(t)} \left[ 1 - \frac{iU_t}{2\omega} \right] \psi_S \; ; \; H_y = ik_\perp \psi_S \; ; \; H_z = -ik_{II}\psi_S \tag{26}$$

where we see that, due to the derivative of $U$ in factor of $i$, the electric field and the vector potential (which is here also the convenient generating function) are not in quadrature. One also sees that the fields do not depend sinusoidally of $t$, so that some significant reshaping of the wave in the medium will occur[5]. Rather than going into such details, we would like to show some important physical effects occurring in the reflection of light on such a non stationary media. Once again the traditional method of fields continuity at the interfaces can be used, and one so obtains the laws of reflection of light on such a medium as a combination of a "generalized Descartes law" :

$$\sin \gamma = n_{S,P} \sin \beta \tag{27}$$

where $n_{S,P}$ is a complex refractive index

$$n_S = n_0 U(t) N_S \left(1 - iU_t/2\omega\right)^{-1}$$
$$n_P = n_0 U(t) N_P \left(1 + iU_t/2\omega\right)^{-1} \tag{28}$$

$S$ and $P$ referring to the incident wave (and corresponding respectively to the cases of a TE/TM waves in the medium mentioned above). The incidence angle $\gamma$ being real, it follows that the refraction angle $\beta$ is complex, a situation which is not uncommon since it is encountered in the optics of metals[6] and will have similar consequences (inhomogeneous wave inside the medium with different equal intensity and equal phase planes). However here, the origin of this quite different here : it arises because of the non stationarity, no matter whether the dielectric constant is positive or negative. Also note that the refractive index is here quite different from the index characterizing the dispersion in the medium. The "generalized Fresnel laws" can also be obtained as

$$R_S = \frac{\cos \gamma - \sqrt{n_S^2 - \sin^2 \gamma}}{\cos \gamma + \sqrt{n_S^2 - \sin^2 \gamma}} \tag{29}$$

$$R_P = \frac{n(t) \cos \gamma - \sqrt{n_P^2 - \sin^2 \gamma}}{n(t) \cos \gamma + \sqrt{n_P^2 - \sin^2 \gamma}} \tag{30}$$

It is quite instructing to consider the very simple case of a linear $U$ dependence (which happens to be valid for both incident polarizations) like

$$U(t) = 1 - t/t_0 \quad ; \quad t < t_0 \tag{31}$$

even though it is too limited for many experimental situations, since it allows easy calculations. Writing the instantaneous value of the material optical index (in this framework, the word "refractive is inadequate) $n(t)=n_0 U(t)$, and restricting ourselves to the case of a normal incidence, one obtains for the reflection coefficient ($R_P$ and $R_S$ coincide in this case) :

$$Re(R_S) = \frac{N_S \left(1 - n^2(t)\right)}{N_S \left(1 + n^2(t)\right) + 2n(t)}$$

$$Im(R_S) = \frac{-2in(t)}{\varphi_0 \left[N_S \left(1 + n^2(t)\right) + 2n(t)\right]} \tag{32}$$

which we should compare to the values that one would obtain from a "quasi-stationary" approximation, that is applying the normal Fresnel laws using the instantaneous value of the optical index $n(t)$ :

$$R_S = \frac{1-n}{1+n} = \frac{1-n_0(1-t/t_0)}{1+n_0(1-t/t_0)} \tag{33}$$

which is essentially different from expressions (32), to begin with through the fact that this quasi-stationary value has no imaginary part, contrary to the real one. One thus does an essential mistake by applying such an approximation, which can be traced back to the fact that one neglects the non-stationarity induced dispersion (implicitly setting $U_t=0$, even though one could think that the time dependence of $n$ is accounted for). This is illustrated on figure 3 showing the behavior of these different quantities.

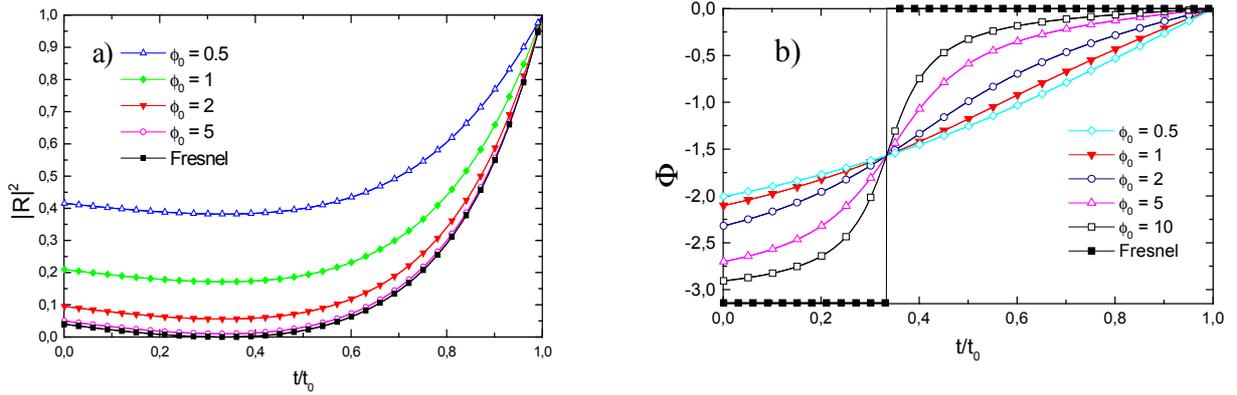

Figure 3 : Reflection coefficient of a non stationary dielectric corresponding to $\varepsilon$ variations of eq (31), for different values of the parameter $\phi_0=2\omega t_0$, compared to the result of a "quasistatic" approximation using the instantaneous value of the standard Fresnel coefficient (a) Intensity reflection coefficient (b) phase. The beam is at normal incidence

From figure 3, one deduces that if for the intensity reflection coefficient, the differences are minor as soon as the time constants for the dielectric constant variations exceed a few laser periods, the phase appears to be much more sensitive to such effects (and therefore constitutes a much better test). Other effects in the reflected fields can be calculated simply from the above method, such as for instance frequency shifts in the reflected field arising from a time dependent reflection phase shift[5].

As a conclusion to this part, we note that for the optics of non-stationary media, very specific effects occur that can be calculated and, probably more important, understood using our "phase coordinate method". In many respects the space-dependent and time-dependent cases appear to be similar.

## 4. Application to non-linear optics

We illustrate in this part how a similar method can be used to study the interaction of a linearly polarized EM transient with an isotropic nonlinear dielectric. Assuming the nonlinearity relaxation time to be shorter than the transient's duration, one can represent the electric displacement $D$ as a continuous function of the electric field $E$: $D = D(E)$. We consider a linearly polarized EM field ($E$ along the $x$ direction, $H$ along the $y$ direction, propagation in the $z$ direction, the indexes being omitted hereafter for lightness in the notations). The Maxwell equations for the field travelling in this medium (assumed to be non-magnetic) can then be written as:

$$\frac{\partial E}{\partial z} = -\frac{1}{c}\frac{\partial H}{\partial t} \quad ; \quad -\frac{\partial H}{\partial z} = \frac{1}{c}\frac{\partial D}{\partial t} \tag{34}$$

We will rewrite this system using the hodograph transform[7] and treating the functions $E$ and $H$ as new independent variables and the variables $z$ and $t$ as new unknown functions $z = z(E,H)$, $t = t(E,H)$. The fundamental system (34) then becomes

$$\frac{\partial t}{\partial H} = -\frac{1}{c}\frac{\partial z}{\partial E} \quad ; \quad \frac{\partial t}{\partial E} = -\frac{1}{c}\frac{\partial D}{\partial E}\frac{\partial z}{\partial H} \tag{35}$$

It is remarkable that the system (35), unlike the system (34), is linear, and the function $\partial D/\partial E$, connected with the nonlinear response of the medium, may be considered as some inhomogeneity in the $(E,H)$ space. Thus, the system (35) is formally analogous to the Maxwell equations with a coordinate – dependent velocity. The system (35) can be reduced to one equation by two approaches, related respectively to two different models of nonlinearity. The first approach is based on the expression of quantities $t$ and $z$ via some generating function $F$

$$t = \frac{1}{c}\frac{\partial F}{\partial E} \quad ; \quad z = -\frac{\partial F}{\partial H} \tag{36}$$

Using (36) one can transform the first equation in the system (35) into an identity, meanwhile the function $F$ is governed by the second equation. Examining the media with $\partial D/\partial E > 0$, and putting thus

$$\frac{\partial D}{\partial E} = n_0^2 U^2(E) \quad ; \quad D = n_0^2 \int_0^E U^2(E)\,dE \tag{37}$$

where $n_0$ is the linear value of refractive index, one can present the equation governing the function $F$ in a dimensionless form (with $E = E_c x$, $H = E_c n_0 h$):

$$\frac{\partial^2 F}{\partial x^2} - U^2(x)\frac{\partial^2 F}{\partial h^2} = 0 \tag{38}$$

Considering the case $U = (1 + E/E_c)^{-2} = (1+x)^{-2}$, introducing a new function $f$ and a new variable $\varphi$, one can rewrite eq. (38) as

$$F = \frac{f}{\sqrt{U}}; \;\varphi = \frac{x}{1+x}; \quad \frac{\partial^2 f}{\partial \varphi^2} - \frac{\partial^2 f}{\partial h^2} = 0 \tag{39}$$

This simple wave equation describes the EM field propagation in a medium with the nonlinearity, defined due to substitution of function $U = (1 + E/E_c)^{-2}$, used above, into eq. (4):

$$D = \frac{n_0^2 E_c}{3}\left[1 - \frac{1}{\left(1 + E/E_c\right)^3}\right] \tag{40}$$

The second approach for transforming the pair (35) to one equation is based, unlike (36), on the presentation of generating function $F$ by means of correlations

$$t = \frac{n_0}{c}\frac{\partial F}{\partial H} \quad ; \quad z = -\frac{1}{n_0 U^2}\frac{\partial F}{\partial E} \tag{41}$$

Introducing the new function $f = F/\sqrt{U}$, using the variable $\varphi = (1 + E/E_c)^{1/3} - 1$ and supposing in this case $U = (1+\varphi)^{-2}$, one can see, that the function $f$ is governed by the same equation (39). The electric displacement $D(E)$, related to this model of $U$, being

$$D = 3n_0^2 E_c \left[1 - \left(1 + \frac{E}{E_c}\right)^{-1/3}\right] \qquad (42)$$

Formulae (40) and (42) present models related respectively to "fast" and "slow" saturation of the nonlinearity. In the limit of a weak pulse field ($E \ll E_c$), one obtains from both (40) and (42) the linear limit: $D = n_0^2 E$. As in the cases considered in sections 2 and 3, we have, through a set of transformations cast our problem in a form admitting simple analytical solutions.

## 5. Conclusions

In this paper, we have shown that the problem of the optics of inhomogeneous or non-stationary media could be formally simplified by moving into a space where phase accumulation recovers linearity with either space of time (the "phase coordinate method"). Under some restrictions concerning the analytical forms used to represent the variations of the dielectric constant, which do not however preclude the possibility of representing all types of variations, one is then able to obtain complete analytical solutions with the essential advantage that one imposes no restriction concerning how fast and how large the variations are, and that one keeps a real physical insight into the optics of such materials. We found that the dispersion laws that are plasma- (or waveguide-) like, the dispersion being essentially determined by the spatial or temporal characteristic scales of the variations ("inhomogeneity or non-stationarity induced dispersion"). We showed that some important consequences can derive from this special type of dispersion, e.g. on the reflection properties of such media, and were able to evaluate at which stage, e.g., quasi-stationary approximations become dangerous and why. Finally we showed that such a method can be applied to other problems, in particular indicating how, with the help of a hodograph transform, the case of an arbitrarily high non-linearity could be approached.

It is also worth mentioning that optics is not the only field in which such methods could be applied. The Schroedinger equation would for instance be relevant of the same methods, yielding solutions to many unsolved problems (or at least solved in the framework of disputable approximations) such as that of the "split-potential".